\documentclass[9pt,twocolumn,twoside]{opticajnl}
\journal{opticajournal} 

\setboolean{shortarticle}{true}

\usepackage{lineno}
\usepackage{afterpage}
\usepackage{placeins}
\usepackage{indentfirst}
\usepackage{caption}
\usepackage{graphicx}
\usepackage{amsmath}
\usepackage{subfig}
\usepackage{subfloat}
\title{Quantum delayed "choice" based on vectorially structured photon} 

\author[1]{Ye Yang}
\author[1]{Shuya Zhang}
\author[1]{Yongkun Zhou}
\author[1]{Xinji Zeng}
\author[1]{Kaixuan Ren}
\author[1]{Dong Wei}
\author[1, *]{Chengyuan Wang}
\author[2, **]{Yun Chen}
\author[1]{Hong Gao}
\author[1]{Fuli Li}

\affil[1]{MOE Key Laboratory for Nonequilibrium Synthesis and Modulation of Condensed Matter, Shaanxi
Province Key Laboratory of Quantum Information and Quantum Optoelectronic Devices, School of Physics,
Xi'an Jiaotong University, 710049, China}
\affil[2]{Department of Physics, Huzhou University, Huzhou 313000, China}

\affil[*]{wcy1992@xjtu.edu.cn}
\affil[**]{1484654436@qq.com}

\begin{abstract}
Whether a photon exhibits wavelike or particlelike behaviour depends on the observation method, as clearly demonstrated by Wheeler's delayed choice (DC) experiments. A key aspect of such experiments is the random determination of the observation device's status, typically controlled by a random number generator or a quantum-controlling apparatus. Here, we propose a novel version of the quantum delayed choice (QDC) experiment by tailoring the quantum state of the single photon into an arbitrary polarization superposition. In this experiment, the "choice" can be considered as being made by the photon's state itself at the moment of observation, thereby violating classical causality. Additionally, we observe the morphing behaviour of the single photon between wavelike and particlelike characteristics, which challenges the classical picture of waves and particles. Utilizing the quantum state of the photon rather than the quantum-controlling devices not only facilitates the implementation of the QDC experiment but also helps deepen the understanding of Bohr's complementarity principle.

\end{abstract}

\setboolean{displaycopyright}{false} 

\begin{document}

\maketitle

\par
\begin{figure}[h]
\centering
\includegraphics[width=0.8\linewidth]{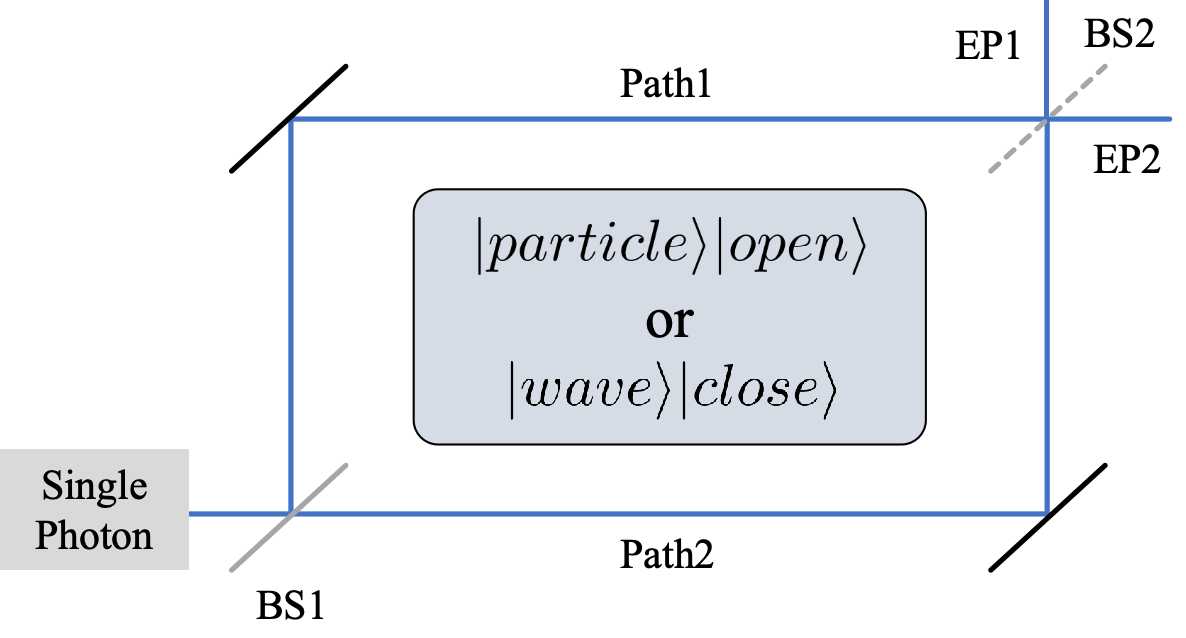}
\caption{Mach-Zender Interferometer (MZI) based schematic diagram of the theoretical model. Where $\theta$ denotes the relative phase between two paths. BS, 50:50 beam splitter; EP, exit port.}
\label{MZI model}
\end{figure}

 The wave-particle duality of light is a fundamental concept in quantum physics that challenges classical intuition\cite{RN305}. According to Bohr's complementarity principle\cite{RN440}, whether a photon exhibits wavelike or particle-like behaviour depends on the method of observation. A classical setup used to investigate this phenomenon is the Mach-Zehnder interferometer (MZI). As illustrated in Fig.\ref{MZI model}, in the absence of beam splitter 2 (BS2), detectors at the two exit ports (EP1 and EP2) can determine which path the single photon has passed through, resulting in particle-like behaviour. Conversely, when BS2 is present, the which-path information is erased\cite{RNoamErasure,RN310,RN309}, and the photon exhibits wavelike behaviour. This experimental design effectively demonstrates the wave-particle duality of photons and is consistent with Bohr's complementarity principle\cite{RN297,RN367,RN359,RN540}.

\par However, the hidden variable theory\cite{RN557,RN373} suggests that the photon may acquire information of subsequent observations setups before entering the MZI, thereby exhibiting either wavelike or particlelike properties in advance. To refute this theory, the delayed choice (DC) experiment was proposed\cite{RN390,RN357,RN439,RN559,RN560}. A core feature of the DC experiment is the random selection of the observation method after the photon has entered the MZI. Specifically, a random decision is made on whether to insert BS2 after the photon has passed through the BS1. This design ensures that the photon is unable to predict the subsequent observational setup, thereby preventing the premature determination of its behaviour according to the hidden variable. This proposal has been realised in various experimental systems, including single-photon\cite{RN296,RN311,RN361}, biphoton\cite{RN386,RN371,RN302}, thermal light\cite{RN304}, neutral atoms\cite{RN312,RN362}, and electrons\cite{RN381}. Further, in 2011, a quantum version of the delayed choice (QDC) framework was proposed \cite{RN313}, in which the choice is made by a quantum-controlled BS2 rather than by the classical control typically used in DC experiments. That is, before the photon enters the MZI, BS2 exists in a quantum superposition state of being inserted and uninserted, which correspond to the closed and open states of the interferometer, respectively, given by $(|\text{close}\rangle+|\text{open}\rangle)/\sqrt{2}$. Based on this, the possibility of the photon having prior knowledge of the measurement device is eliminated in physical principle\cite{RN313}.
  \begin{figure*}[t]
  \centering
\includegraphics[width=0.8\linewidth]{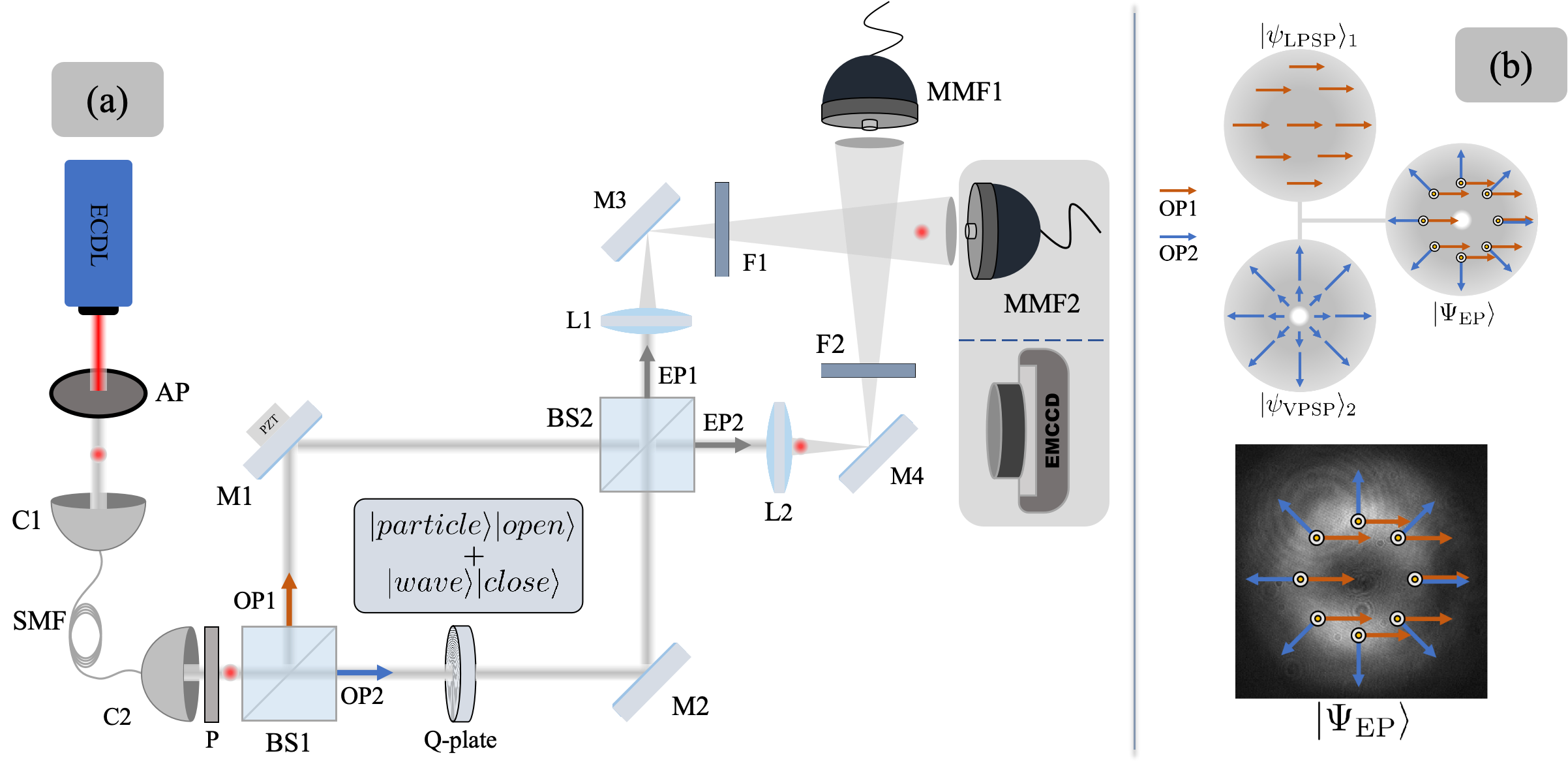}
\caption{The schematic diagram of a quantum delay selection experimental system utilizing VPSP. ECDL, External-Cavity Diode Laser; AP, Attenuator Plate; SMF, Single Mode Fibre; C, Coupler; P, Polariser; BS, 50:50 Beam Splitter; M, Mirror; Q-plate, Variable spiral plate, the topological charge $q=1/2$; PZT, Piezoelectric Transducer; L, Lens ($f=50mm$); F, 795nm optical filter; MMF, Multi-mode fibre (combination of coupler and multimode fibre); EMCCD, Electron-Multiplying Charge-coupled Device. \textbf{(a)} Coherent light generated from the ECDL is attenuated to a single photon level by the AP. C1 and C2 are connected to an SMF that purifies the photon to a LPSP state. After through P, polarization is polarized to $|H\rangle$. The photon then enters the Mach-Zehnder Interferometer (MZI), which is constructed from M1 (with PZT), M2, BS1, and BS2. The PZT integrated onto M1 enables sweeping the relative phase $\theta$ of the interferometer, while the Q-plate modulates the photon into VPSP state. After BS2 combines the optical path, L1 and L2 focus the light field onto the surfaces of M3 and M4. This arrangement allows for adjust the position of the light field and facilitating beam expansion. Subsequently, the light field is further filtered from noise by F1 and F2 before being directed onto the detector which is connected with MMF1 and MMF2. And the detector of EP2 could be replaced by EMCCD. \textbf{(b)} The superposition of $|\psi_{LPSP}\rangle_{1}$ and $|\psi_{\text{VPSP}}\rangle_{2}$. The VPSP discussed here has a radial distribution of polarization. There is a phase singularity in the centre and the amplitude is 0. And the LPSP light has a horizontal polarisation. Following the superposition performed by the interferometer, the state $|\Psi_{\text{EP}}\rangle$ can be obtained.}
\label{experimentSystem}
\end{figure*}

\par In this paper, we propose a novel QDC scheme that utilizes vector-polarized\cite{RN564,RN565,RN566,RN567,RN568,RN551} single photon (VPSP), whose polarization patterns are inhomogeneously distributed across the spatial mode. The VPSP are generated through Q-plate (QP) modulation of a linearly polarized single photon (LPSP), and can be considered in a quantum superposition of various polarization states. In this scheme, path 2 of the MZI carries a VPSP state, while path 1 carries a LPSP state. Upon merging at BS2, if the VPSP state in path 2 "collapses" into the same polarization as the LPSP state in path 1, the single photon will interfere and exhibit wavelike behaviour. Conversely, if the VPSP state "collapses" into a polarization orthogonal to that of the LPSP state, the photons will not interfere and will exhibit particlelike behaviour. Benefitting from the easily accessible and plentiful polarization properties of VPSP, we are able to observe continuous morphing between wave and particle behaviour within a single experimental setup. Our scheme helps a deeper understanding of complementarity in quantum mechanics.

\par The experimental setup is schematically displayed in Fig.\ref{experimentSystem}a. An LPSP is obtained by the attenuation of a coherent light generated by an External-Cavity Diode Laser (ECDL), which then is incidented into an MZI. The MZI consists of two BSs (BS1 and BS2) and two mirrors (M1 and M2). A PZT attached to M1 can sweep the relative phase of the two optical paths (OP1 and OP2). A Q-Plate (topological charge $q=0.5$) is inserted into OP2, allowing the manipulation of LPSP into VPSP \cite{RN461,RN462}, whose polarization distribution is illustrated in Fig.\ref{experimentSystem}b. Such a transformation process can be expressed as:
\begin{equation}
|\psi_{\text{VPSP}}(\mathbf{r}, t)\rangle=\hat{\text{QP}}|\psi_{\text{LPSP}}^{H}(r, t)\rangle_{2}=e^{-i\omega t}A_{2}(r)|P(\varphi)\rangle
\end{equation}
where $\hat{\text{QP}}$ denotes the operator of QP that acts on the photon's state, $\psi_{\text{LPSP}}^{H}$ represents photon in horizontal polarised LPSP state, $A(r)$ represents the amplitude profile that generally depends on the radial coordinate $r$, and the set $\{|P(\varphi)\rangle=sin \varphi |H\rangle+cos \varphi |V\rangle \} $ comprises the transverse unit vectors specifying the local mode polarization as a function of the azimuthal angle $\varphi$. Additionally, $\omega$ denotes the frequency of the light field. When a single photon passes through the MZI, the quantum state in OP1 is represented by the LPSP state $|\psi_{\text{LPSP}}^{H}(r, t)\rangle_{1}=e^{-i\omega t}A_{1}(r)|H\rangle$, and the quantum state in OP2 is represented by $|\psi_{\text{VPSP}}(\mathbf{r}, t)\rangle_{2}$. Consequently, the transverse modes of the light fields at EP1, corresponding to the upper exit ports of BS2, can be described as follows:
\begin{equation}
    |\psi_{\text{EP1}}(\textbf{r}, t)\rangle=\frac{1}{\sqrt{2}}[e^{i\theta}|\psi_{\text{LPSP}}^{H}(r, t)\rangle_{1}+|\psi_{\text{VPSP}}(\textbf{r}, t)\rangle_{2}]
    \label{eq:ep1-PolarisationState}
\end{equation}
In this context, the subscripts '1' and '2' refer to OP1 and OP2, respectively. The term $e^{i\theta}$ represents the relative phases between the two paths induced by PZT on M1. This description is same applies to $|\psi_{\text{EP2}}(\textbf{r}, t)\rangle$, but with an additional relative phase of $\pi$ introduced by a further reflection. At EP1 (and similarly at EP2), the state can be obtained as the integral of the quantum states over all points, as described below:

\begin{equation}
    |\Psi_{\text{EP1}}\rangle=\int C(\varphi)r|\psi_{\text{EP1}}(\textbf{r}, t)\rangle \text{d}\varphi \text{d}r
    \label{eq:ep1-integralState}
\end{equation}
Where $C(\varphi)$ represents the continuous components of quantum states that depend on $\varphi$. For simplicity and without loss of generality, we could ignore the radial distribution, as the polarization variation depends solely on $\varphi$ (see Fig.\ref{experimentSystem}b)\cite{RN505}. Furthermore, we assume that $A_{1}(r)\approx A_{2}(r)$ for  $ r\neq 0$.

\par In this scenario, the indistinguishability of the photon's path information is contingent upon the transverse mode of the light field. As shown in Fig.\ref{experimentSystem}b, when a photon exits the MZI, its path information is preserved due to the orthogonality of the polarization states associated with the two paths. Conversely, when the polarizations of the two paths are identical, the photon's path information is erased. In previous studies, this delayed-choice scenario can be implemented using polarization modulation devices\cite{RN387,RN368}. In this study, while the photon is prepared in the quantum state $|\Psi_{\text{EP1}}\rangle$, the intrinsic properties of this state directly facilitate the delayed-choice experiment. When a photon exits the MZI, its polarization-path information exists in an arbitrary angle superposition, according to Eq.\ref{eq:ep1-PolarisationState}, Eq.\ref{eq:ep1-integralState} and assume $A_{1}(r)\approx A_{2}(r)$, which can be described as:
\begin{equation}
    |\Psi_{\text{EP1}}\rangle=\frac{1}{\sqrt{2}}\int C(\varphi)r e^{-i\omega t}A_{1}(r)[e^{i\theta}|H\rangle_{1}+|P(\varphi)\rangle_{2}] \text{d}\varphi \text{d}r
    \label{eq:ep1-polarisationSuperpositon}
\end{equation}
Until detection by the EMCCD (Electron-Multiplying Charge-coupled Device), the choice of whether to erase the path information is instantaneously determined, transitioning from an integral form quantum superposition state to a specific state $|\psi_{\text{EP1}}(\textbf{r}', t')\rangle\propto|\psi'\rangle=[e^{i\theta}|H\rangle_{1}+|P(\varphi')\rangle_{2}]$, in which the state $|P(\varphi')\rangle_{2}$ determines the state of the observation setup. This implies that if $|P(\varphi')\rangle_{2}=|V\rangle_{2}$ (or $|P(\varphi')\rangle_{2}=|H\rangle_{2}$), the path information is retained (or erased) and the photon behaves like a particle (or wave). Furthermore, due to the integral superposition of polarization states, the observation setup is also in a integral superposition state. Consequently, the state of the photon and the setup $|\psi'\rangle$ becomes\cite{RN313}:
\begin{equation}
|\psi'\rangle=C_1(\varphi')|particle\rangle|open\rangle+C_2(\varphi')|wave\rangle|close\rangle
    \label{eq:superposition of waveAndParticle}
\end{equation}
In which $|particle\rangle=\frac{1}{\sqrt{2}}(e^{i\theta}|H\rangle+|V\rangle), |wave\rangle=\frac{1}{\sqrt{2}}(e^{i\theta}+1)|H\rangle$ and $C_1(\varphi)=\langle particle|\psi\rangle, C_2(\varphi)=\langle wave|\psi\rangle$. The term "open" indicates that path information is retained, while "close" indicates that path information is erased; this is equivalent to the open and closed states of the MZI, respectively. 

\begin{figure}[t]
    \centering
    \subfloat(a){\includegraphics[width=0.45\linewidth]{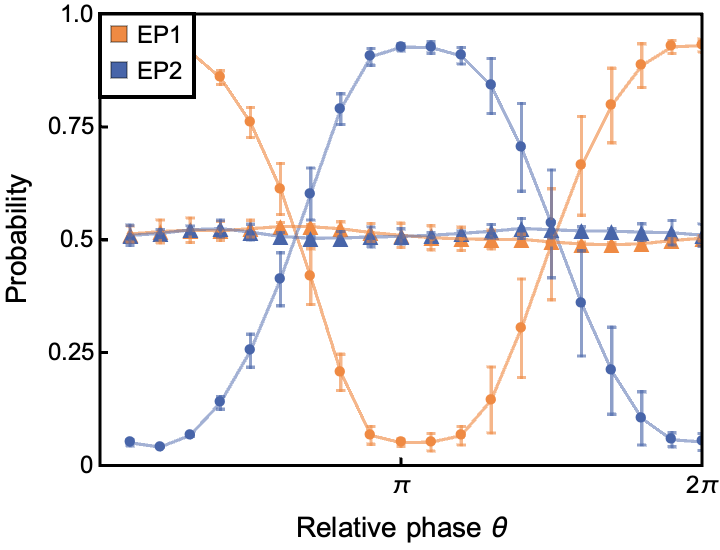}}
    \hfill
    \subfloat(b){\includegraphics[width=0.45\linewidth]{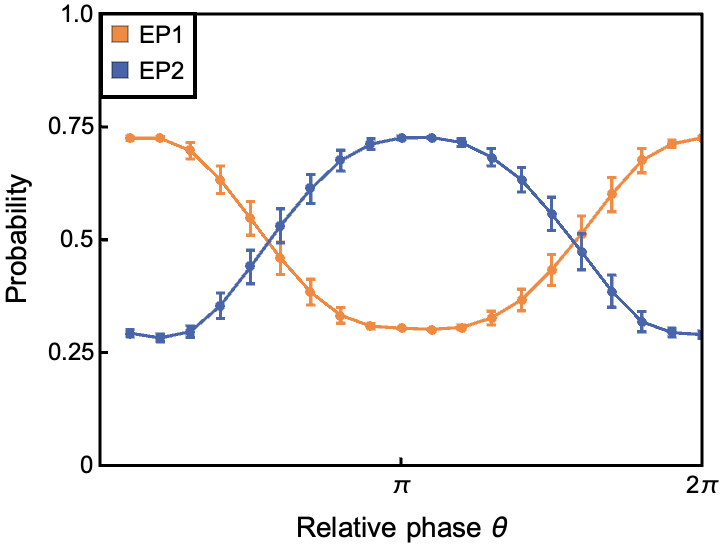}}
    \caption{The behaviour of photon states projected at different points. The orange and blue markers correspond to EP1 and EP2, respectively, and represent the probability of detecting a single photon while the MZI at the relative phase of $\theta$. \textbf{(a)} Interference measurements in which the polarization-path information of a single photon is either retained ($\varphi=0$, marked by triangles) or erased ($\varphi=\pi/2$, marked by dots) through delayed observation. \textbf{(b)} Measurements in which the polarization path information is partially erased ($\varphi=\pi/4$) leave the photon in a superposition of wavelike and particlelike states $|\psi\rangle=C_1(\pi/4)|particle\rangle|open\rangle+C_2(\pi/4)|wave\rangle|close\rangle$.}
    \label{interferenceCurve}
\end{figure}

\par To verify the our concept experimentally, the experimental setup is shown in Fig.\ref{experimentSystem}a. A photon is prepared in a quantum state $|\Psi_{\text{EP}}\rangle$ and export from the interferometer. The light field is focused onto the surfaces of mirrors M3 and M4 using lenses L1 and L2. By adjusting M3 and M4, the transverse mode of the single photon is expanded and projected onto MMF1 and MMF2 (a combination of a coupler and multimode fibre). To ensure the single-photon nature of the system, EP1 and EP2 are measured using coincidence detection, with the coincidence rate being $1.83\times10^{-3}$ times the counting rate of EP1 (or EP2). This indicates that, in most cases, the system contains only a single photon, thereby excluding multi-photon interference effects. The single photon is subsequently observed at various transverse positions (see Fig.\ref{experimentSystem}b), displaying the behaviour illustrated in Fig.\ref{interferenceCurve}. When $\varphi = \pi/2$, the path information is erased due to the indistinguishability of the polarization states, resulting in wavelike behaviour (Fig.\ref{interferenceCurve}a, marked by dots). Conversely, when $\varphi = 0$, the path information is retained due to the distinguishability of the polarization states, leading to particlelike behaviour (Fig.\ref{interferenceCurve}a, marked by triangles). When $\varphi = \pi/4$, the path information is partially erased due to the partial distinguishability of the polarization states, and the single photon exhibits a mixed wavelike behaviour (Fig.\ref{interferenceCurve}b). The interference fringe visibility of the single-photon distribution is given by $V(\varphi)=\frac{I_{\text{max}}(\varphi)-I_{\text{min}}(\varphi)}{I_{\text{max}}(\varphi)+I_{\text{min}}(\varphi)}$, with values $V(0)=0.02\pm0.001$, $V(\pi/4)=0.44\pm0.01$, and $V(\pi/2)=0.91\pm0.02$. 

\begin{figure}
    \centering
    \includegraphics[width=1\linewidth]{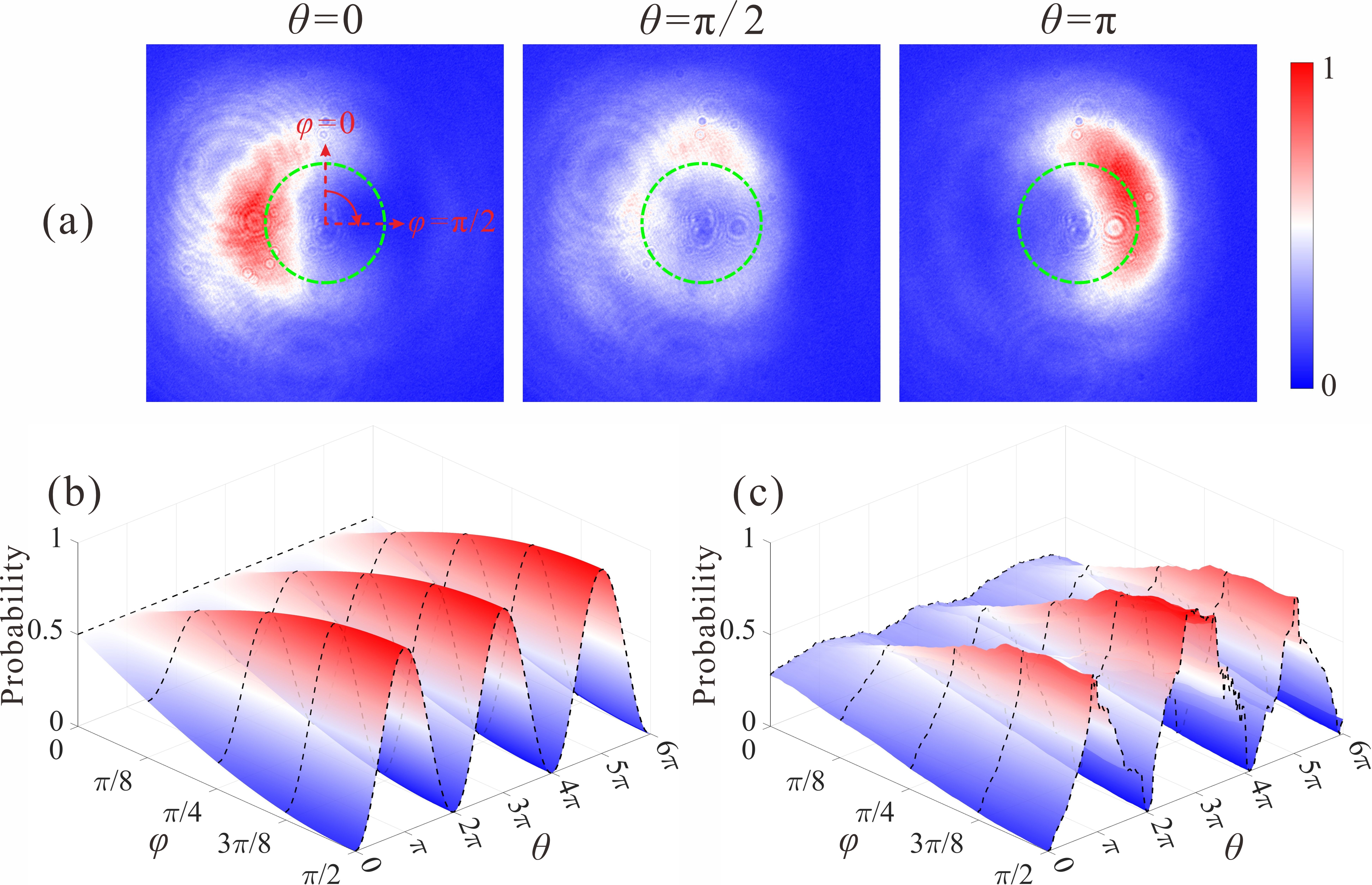}
    \caption{The morphing behaviour between particle ($\varphi=0$) and wave ($\varphi=\pi/2$) of the single photon and its photon number accumulation pattern. \textbf{(a)} The pattern of photon number accumulation at various phases $\theta$ as captured by the EMCCD. \textbf{(b)}Theoretically expected morphing behaviour of the single photon. \textbf{(c)} Experimentally observed morphing behaviour of the single photon, obtained by scanning the azimuthal angle $\varphi$ of the ring (green) in (a) and the phase $\theta$ of the MZI.}
    \label{wholeInterGraph}
\end{figure}

\par The above result is implemented using MMFs, which enables the measurement of a specific point in the transverse mode of a single photon. To provide a more intuitive demonstration of the quantum delayed "choice" of the single photon and its morphing behaviour, MMF2 was replaced by an EMCCD, which can implement detection of the entire transverse mode of a single photon. By sweeping the phase $\theta$ of the MZI, the interference behaviour in the entire transverse mode was captured. As shown in Fig.\ref{wholeInterGraph}a, three typical single-photon number accumulation patterns are presented. By scanning the azimuthal angle $\varphi$ of the ring (green) in Fig.\ref{wholeInterGraph}a and the phase $\theta$ of the MZI, the single-photon morphing behaviour was demonstrated, as shown in Fig.\ref{wholeInterGraph}c. If the single photon eventually "collapses" at $\varphi = 0$, its population distribution becomes independent of the relative phase $\theta$, indicating particle-like behaviour. With $\varphi$ gradually transitioning, it can be seen that the behaviour of photons is gradually transitioning from particlelike ($\varphi=0$) to wavelike ($\varphi=\pi/2$), in good agreement with the theoretical expectations shown in Fig.\ref{wholeInterGraph}b. This challenges the classical view of the photon as either a particle or a wave. Fig.\ref{wholeInterGraph}c intuitively illustrates that the wavelike and particlelike behaviours can be regarded as two limiting cases of the morphing behaviour. In contrast to the DC experiment, which is controlled by a random number generator together with classical device, the "choice" whether to erase the photon’s path information in this experiment occurs at the moment of detection. In other words, the final "choice" can be considered as being made by the photon's state itself. This demonstrates the validity of the delayed choice principle and challenges the classical conception of cause and effect.
\par In conclusion, we have experimentally implemented a novel QDC scheme using VPSP, which is in a quantum superposition of all radial polarizations. Our scheme comprises of an attenuated single photon source carrying linear polarization and a typical MZI, with a Q-plate inserted in one arm of the MZI to generate the VPSP. When the VPSP and LPSP in two arms of the MZI converge at BS, if the VPSP "collapses" into the same polarization as the LPSP, the single photon will interfere and exhibit wavelike behaviour. On the contrary, if the VPSP "collapses" into a polarization orthogonal to that of the LPSP, the photon will not interfere and display particlelike behaviour. By taking advantage of the VPSP containing a quantum superposition of all radial polarizations, we are able to observe continuous morphing between wave and particle behaviour within a single experimental setup. Distinctive from other DC experiments that rely on random number generators to control the post-measurement method for excluding the hidden variable, we leverage the quantum property of the VPSP’s polarization behaviour to demonstrate the QDC experiment in a device-independent way. This approach is more easily accessible and offers a more intuitive insight into the wave-particle duality.

\section*{Funding.}This work is supported by the National Natural Science Foundation of China (NSFC) (12104358, 12104361, 12304406 and 12175168) and the Shaanxi Fundamental Science Research Project for Mathematics and Physics (22JSZ004, and 23JSQ014).
\bibliography{reference}

\bibliographyfullrefs{reference}


\ifthenelse{\equal{\journalref}{aop}}{%
\section*{Author Biographies}
\begingroup
\setlength\intextsep{0pt}
\begin{minipage}[t][6.3cm][t]{1.0\textwidth} 
  \begin{wrapfigure}{L}{0.25\textwidth}
    \includegraphics[width=0.25\textwidth]{john_smith.eps}
  \end{wrapfigure}
  \noindent
  {\bfseries John Smith} received his BSc (Mathematics) in 2000 from The University of Maryland. His research interests include lasers and optics.
\end{minipage}
\begin{minipage}{1.0\textwidth}
  \begin{wrapfigure}{L}{0.25\textwidth}
    \includegraphics[width=0.25\textwidth]{alice_smith.eps}
  \end{wrapfigure}
  \noindent
  {\bfseries Alice Smith} also received her BSc (Mathematics) in 2000 from The University of Maryland. Her research interests also include lasers and optics.
\end{minipage}
\endgroup
}{}

\end{document}